# Towards a Number Theoretic Discrete Hilbert Transform

Renuka Kandregula

## Abstract

This paper presents an approach for the development of a number theoretic discrete Hilbert transform. The forward transformation has been applied by taking the odd reciprocals that occur in the DHT matrix with respect to a power of 2. Specifically, the expression for a 16-point transform is provided and results of a few representative signals are provided. The inverse transform is the inverse of the forward 16-point matrix. But at this time the inverse transform is not identical to the forward transform and, therefore, our proposed number theoretic transform must be taken as a provisional result.

## Introduction

The Discrete Fourier Transform (DFT) has a number theoretic version that has many applications [1]-[4]. We would like to have a similar number theoretic version of the Discrete Hilbert Transform (DHT). This paper presents a number theoretic DHT although it does not represent a general solution.

## The Basic Discrete Hilbert Transform

The basic Discrete Hilbert Transform (DHT) of discrete data f(n) where n = (-∞,…,-1,0,1,…,∞) was given by Kak [5]:

$$DHT\{f(n)\} = g(k) = \begin{cases} \dfrac{2}{\pi} \displaystyle\sum_{n \ odd} \dfrac{f(n)}{k-n}; & k \ even \\ \dfrac{2}{\pi} \displaystyle\sum_{n \ even} \dfrac{f(n)}{k-n}; & k \ odd \end{cases} \qquad (1)$$

The inverse Discrete Hilbert Transform (DHT) is given as:

$$f(n) = \begin{cases} -\dfrac{2}{\pi} \displaystyle\sum_{k \ odd} \dfrac{g(k)}{n-k}; & n \ even \\ -\dfrac{2}{\pi} \displaystyle\sum_{k \ even} \dfrac{g(k)}{n-k}; & n \ odd \end{cases} \qquad (2)$$

The Hilbert transform has many applications in signal processing, imaging, modulation and demodulation, determination of instantaneous frequency and in cryptography.

The discrete Hilbert transform (DHT) has several forms [6]-[8]. In [9]-[10], an application of DHT to data hiding and measure of randomness is given.

## The Matrix Form of the DHT

The matrix form of the DHT requires that the data be of finite length. Since the DHT transform is defined for an infinite number of points, limitation of the DHT transform signal to a finite set would set up an approximation in the signal that is recovered.

The DHT is given below for data n=0, 1, 2, … :

$$
\begin{bmatrix} g(0) \\ g(1) \\ g(2) \\ g(3) \\ g(4) \\ g(5) \\ \cdot \\ \cdot \\ \cdot \end{bmatrix} = \frac{2}{\pi} \begin{bmatrix} 0 & \frac{1}{-1} & 0 & \frac{1}{-3} & 0 & \frac{1}{-5} & 0 & \frac{1}{-7} & \cdot \\ \frac{1}{1} & 0 & \frac{1}{-1} & 0 & \frac{1}{-3} & 0 & \cdot & \cdot & \cdot \\ 0 & \frac{1}{1} & 0 & \frac{1}{-1} & 0 & \frac{1}{-3} & \cdot & \cdot & \cdot \\ \frac{1}{3} & 0 & \frac{1}{1} & 0 & \frac{1}{-1} & 0 & \cdot & \cdot & \cdot \\ 0 & \frac{1}{3} & 0 & \frac{1}{1} & 0 & \frac{1}{-1} & \cdot & \cdot & \cdot \\ \frac{1}{5} & 0 & \frac{1}{3} & 0 & \frac{1}{1} & 0 & \cdot & \cdot & \cdot \\ 0 & \cdot & \cdot & \cdot & \cdot & \cdot & \cdot & \cdot & \cdot \\ \frac{1}{7} & \cdot & \cdot & \cdot & \cdot & \cdot & \cdot & \cdot & \cdot \\ \cdot & \cdot & \cdot & \cdot & \cdot & \cdot & \cdot & \cdot & \cdot \end{bmatrix} \begin{bmatrix} f(0) \\ f(1) \\ f(2) \\ f(3) \\ f(4) \\ f(5) \\ \cdot \\ \cdot \\ \cdot \end{bmatrix}
$$

We see that all the entries in the DHT matrix are odd reciprocals. To construct a number theoretic version of this matrix, we would need to write these reciprocals modulo a suitably chosen number.

In principle, the task of constructing the number theoretic DHT matrix should be easy. In reality, it is a problem since the DHT matrix is defined for an infinite number of index values whereas we are seeking a finite version of the matrix.

In some practical applications we would need to perform the DHT for a finite number of points. In the next section, we present a 16-point number theoretic DHT matrix.

## The 16 point number theoretic DHT matrix

We have obtained a 16 by 16 matrix from the DHT matrix by performing a mod 16 operation on the original DHT matrix values. The reason that the modulus was chosen to be 16 is that all the values in the matrix are odd and therefore there will be a unique value modulo 16. The obtained 16 by 16 matrix is multiplied with the input values to obtain another matrix which we plot it as the transformed image. The obtained transformed values are multiplied by the inverse of the 16 by 16 matrix to regain back the values and are plotted to regain the original image.

The 16-point number theoretic DHT matrix is given below.

$$\begin{bmatrix}
1 & 0 & 11 & 0 & 13 & 0 & 7 & 0 & 9 & 0 & 3 & 0 & 5 & 0 & 15 & 0 \\
0 & 1 & 0 & 11 & 0 & 13 & 0 & 7 & 0 & 9 & 0 & 3 & 0 & 5 & 0 & 15 \\
5 & 0 & 1 & 0 & 11 & 0 & 13 & 0 & 7 & 0 & 9 & 0 & 3 & 0 & 5 & 0 \\
0 & 5 & 0 & 1 & 0 & 11 & 0 & 13 & 0 & 7 & 0 & 9 & 0 & 3 & 0 & 5 \\
3 & 0 & 5 & 0 & 1 & 0 & 11 & 0 & 13 & 0 & 7 & 0 & 9 & 0 & 3 & 0 \\
0 & 3 & 0 & 5 & 0 & 1 & 0 & 11 & 0 & 13 & 0 & 7 & 0 & 9 & 0 & 3 \\
9 & 0 & 3 & 0 & 5 & 0 & 1 & 0 & 11 & 0 & 13 & 0 & 7 & 0 & 9 & 0 \\
0 & 9 & 0 & 3 & 0 & 5 & 0 & 1 & 0 & 11 & 0 & 13 & 0 & 7 & 0 & 9 \\
7 & 0 & 9 & 0 & 3 & 0 & 5 & 0 & 1 & 0 & 11 & 0 & 13 & 0 & 7 & 0 \\
0 & 7 & 0 & 9 & 0 & 3 & 0 & 5 & 0 & 1 & 0 & 11 & 0 & 13 & 0 & 7 \\
13 & 0 & 7 & 0 & 9 & 0 & 3 & 0 & 5 & 0 & 1 & 0 & 11 & 0 & 13 & 0 \\
0 & 13 & 0 & 7 & 0 & 9 & 0 & 3 & 0 & 5 & 0 & 1 & 0 & 11 & 0 & 13 \\
11 & 0 & 13 & 0 & 7 & 0 & 9 & 0 & 3 & 0 & 5 & 0 & 1 & 0 & 11 & 0 \\
0 & 11 & 0 & 13 & 0 & 7 & 0 & 9 & 0 & 3 & 0 & 5 & 0 & 1 & 0 & 11 \\
1 & 0 & 11 & 0 & 13 & 0 & 7 & 0 & 9 & 0 & 3 & 0 & 5 & 0 & 1 & 0 \\
0 & 1 & 0 & 11 & 0 & 13 & 0 & 7 & 0 & 9 & 0 & 3 & 0 & 5 & 0 & 1
\end{bmatrix}$$

It is a circulant matrix just like the standard DHT matrix. It has alternate values of 0, which is what makes the odd and even values of the signal be processed differently.

The Inverse 16 by 16 Matrix is given below. Its structure is similar to the forward transformation but not identical.

$$\begin{bmatrix}
-0.071 & 0.000 & 0.004 & 0.000 & -0.002 & 0.000 & 0.018 & 0.000 & -0.031 & 0.000 & 0.049 & 0.000 & 0.042 & 0.000 & 0.012 & 0.000 \\
0.000 & -0.071 & 0.000 & 0.004 & 0.000 & -0.002 & 0.000 & 0.018 & 0.000 & -0.031 & 0.000 & 0.049 & 0.000 & 0.042 & 0.000 & 0.012 \\
0.000 & 0.000 & -0.060 & 0.000 & 0.004 & 0.000 & -0.002 & 0.000 & 0.018 & 0.000 & -0.031 & 0.000 & 0.049 & 0.000 & 0.042 & 0.000 \\
0.000 & 0.000 & 0.000 & -0.060 & 0.000 & 0.004 & 0.000 & -0.002 & 0.000 & 0.018 & 0.000 & -0.031 & 0.000 & 0.049 & 0.000 & 0.042 \\
0.000 & 0.000 & 0.042 & 0.000 & -0.060 & 0.000 & 0.004 & 0.000 & -0.002 & 0.000 & 0.018 & 0.000 & -0.031 & 0.000 & 0.049 & 0.000 \\
0.000 & 0.000 & 0.000 & 0.042 & 0.000 & -0.060 & 0.000 & 0.004 & 0.000 & -0.002 & 0.000 & 0.018 & 0.000 & -0.031 & 0.000 & 0.049 \\
0.000 & 0.000 & 0.049 & 0.000 & 0.042 & 0.000 & -0.060 & 0.000 & 0.004 & 0.000 & -0.002 & 0.000 & 0.018 & 0.000 & -0.031 & 0.000 \\
0.000 & 0.000 & 0.000 & 0.049 & 0.000 & 0.042 & 0.000 & -0.060 & 0.000 & 0.004 & 0.000 & -0.002 & 0.000 & 0.018 & 0.000 & -0.031 \\
0.000 & 0.000 & -0.031 & 0.000 & 0.049 & 0.000 & 0.042 & 0.000 & -0.060 & 0.000 & 0.004 & 0.000 & -0.002 & 0.000 & 0.018 & 0.000 \\
0.000 & 0.000 & 0.000 & -0.031 & 0.000 & 0.049 & 0.000 & 0.042 & 0.000 & -0.060 & 0.000 & 0.004 & 0.000 & -0.002 & 0.000 & 0.018 \\
0.000 & 0.000 & 0.018 & 0.000 & -0.031 & 0.000 & 0.049 & 0.000 & 0.042 & 0.000 & -0.060 & 0.000 & 0.004 & 0.000 & -0.002 & 0.000 \\
0.000 & 0.000 & 0.000 & 0.018 & 0.000 & -0.031 & 0.000 & 0.049 & 0.000 & 0.042 & 0.000 & -0.060 & 0.000 & 0.004 & 0.000 & -0.002 \\
0.000 & 0.000 & -0.002 & 0.000 & 0.018 & 0.000 & -0.031 & 0.000 & 0.049 & 0.000 & 0.042 & 0.000 & -0.060 & 0.000 & 0.004 & 0.000 \\
0.000 & 0.000 & 0.000 & -0.002 & 0.000 & 0.018 & 0.000 & -0.031 & 0.000 & 0.049 & 0.000 & 0.042 & 0.000 & -0.060 & 0.000 & 0.004 \\
0.071 & 0.000 & 0.000 & 0.000 & 0.000 & 0.000 & 0.000 & 0.000 & 0.000 & 0.000 & 0.000 & 0.000 & 0.000 & 0.000 & -0.071 & 0.000 \\
0.000 & 0.071 & 0.000 & 0.000 & 0.000 & 0.000 & 0.000 & 0.000 & 0.000 & 0.000 & 0.000 & 0.000 & 0.000 & 0.000 & 0.000 & -0.071
\end{bmatrix}$$

## Results and Discussions

The following gives a detailed explanation and examples to support the statements that have been made in the earlier section of the paper. Upon implementing the MATLAB code, we have obtained the graphs that show that the original, transformation of the original; using the 16 by 16 matrix and also the inverse of the transformation using the Inverse matrix. We now provide results by considering various input values.

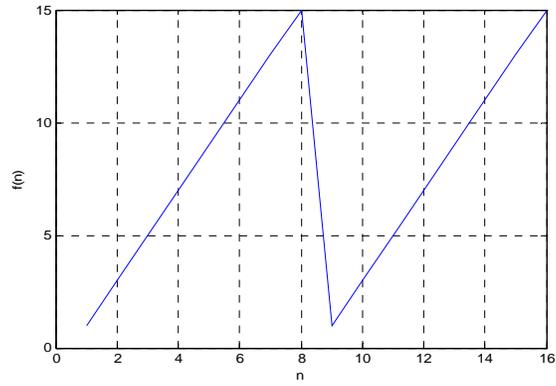

**Fig a**

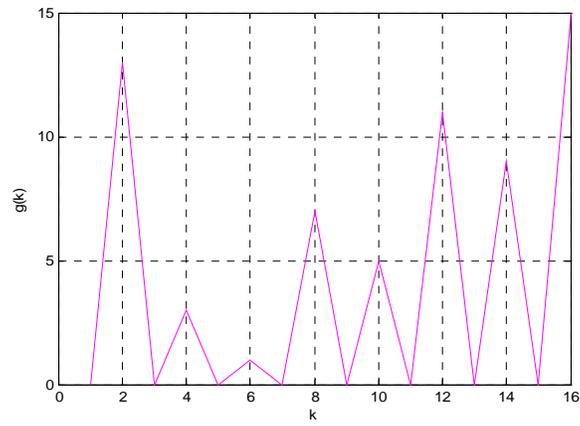

**Fig b**

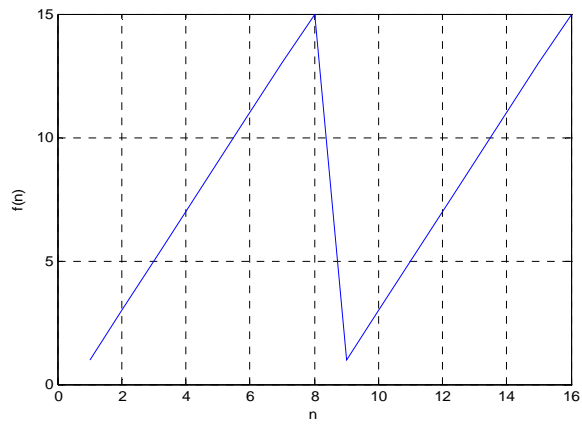

**Fig c**

Figure 1: Input Values = 1   3   5   7   9   11   13   15   13   5   7   9   11   13   15 (a) Original (b) Transformed Image (c) Inverse of the Transformed

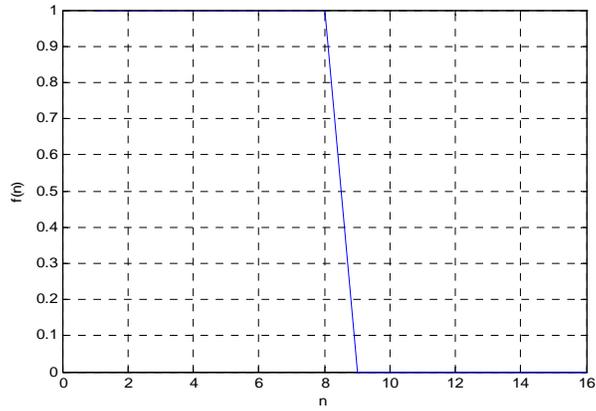

**Fig a**

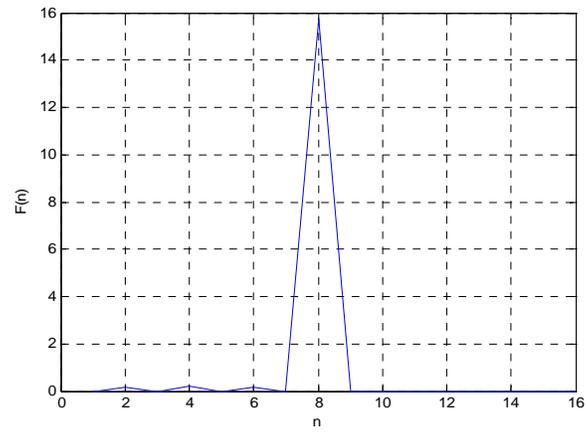

**Fig b**

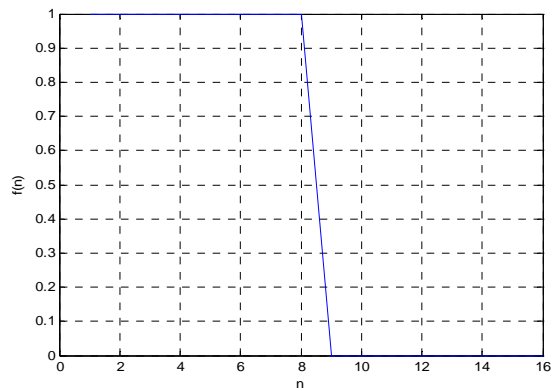

**Fig c**

Figure 2 : Input Values =   1   1   1   1   1   1   1   1   0   0   0   0   0   0   0   0
(a) Original (b) Transformed Image (c) Inverse of the Transformed

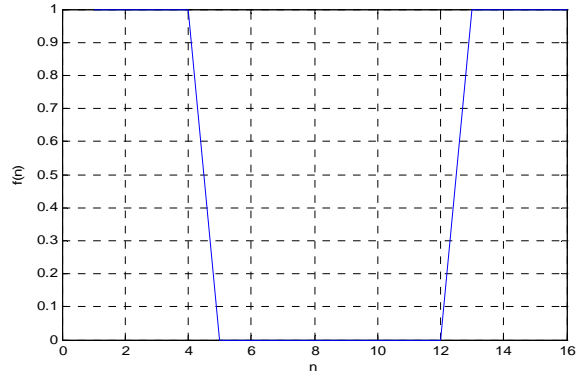

**Fig a**

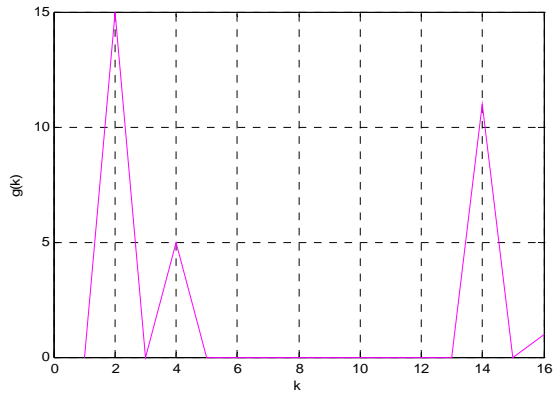

**Fig b**

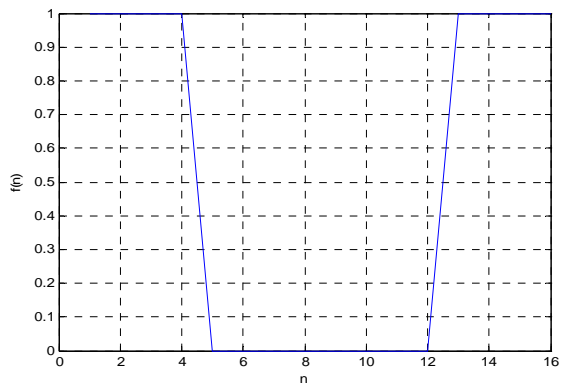

**Fig c**

Figure 3 : Input Values  1   1   1   1   0   0   0   0   0   0   0   0   1   1   1   1
(a) Original (b) Transformed Image (c) Inverse of the Transformed

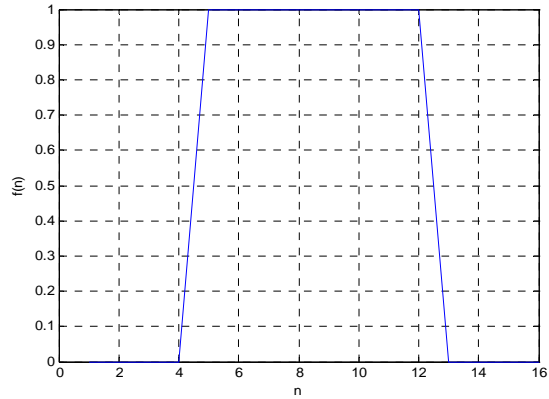

**Fig a**

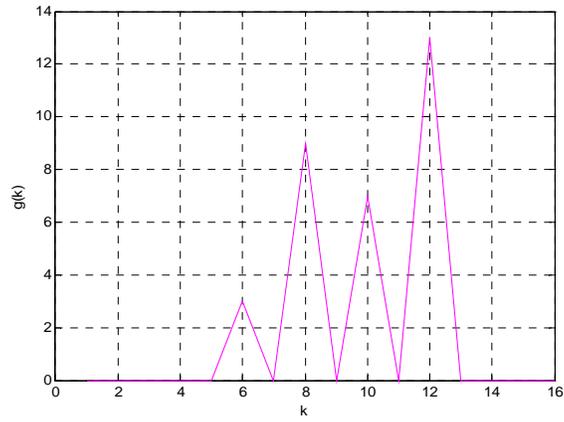

**Fig b**

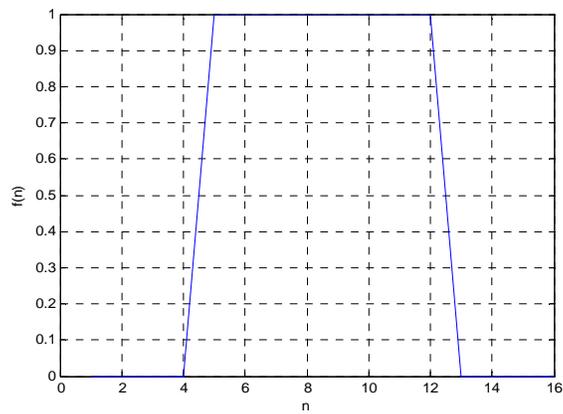

**Fig c**

Figure 4: Input Values  0   0   0   0   1   1   1   1   1   1   1   1   0   0   0   0
Original (b) Transformed Image (c) Inverse of the Transformed

## Conclusions

The number theoretic transform is able to pick out sudden amplitude changes but it doesn't do so neatly. It worked perfectly only if the signal had a single amplitude transition. Specifically, there are several smaller peaks if the signal has more than one amplitude transition.

This work can only be viewed as first step in an approach for the development of a number theoretic discrete Hilbert transform. The forward transformation has been applied by taking the odd reciprocals that occur in the DHT matrix with respect to a power of 2. Specifically, the expression for a 16-point transform is provided and results of a few representative signals are provided. The inverse transform is the inverse of the forward 16-point matrix. But at this time the inverse transform is not identical to the forward transform and, therefore, our proposed number theoretic transform must be taken as a provisional result.

We hope that this work will spur research in the discovery of a true number theoretic discrete Hilbert transform relation.

## References

1. J.M. Pollard, Implementation of Number Theoretic Transforms, Electron Lett., 1976, vol. 12, pp. 378-379.

2. I.S. Reed and K.Y. Liu, A new fast algorithm for computing a complex Number Theoretic Transform, IEEE Trans. Inf. Theory, July 1976.

3. R.C. Agarwal and C.S. Burrus, Number Theoretic Transforms to implement Fast Digital Convolution, Proc. IEEE, vol. 63, No. 4 April, 1975.

4. Liu, K. Y., Reed, I. S., and Truong, T. K., Fast Number-Theoretic Transforms for Digital Filtering, Electron. Lett., 1976, 12, pp. 644-646.

5. S. Kak, The discrete Hilbert transform. Proc. IEEE, vol. 58, pp. 585-586, 1970.

6. S.K. Padala and K.M.M. Prabhu, Systolic arrays for the discrete Hilbert transform. Circuits, Devices and Systems, IEE Proceedings, vol. 144, pp. 259-264, 1997.

7. S. Kak, Hilbert transformation for discrete data. International Journal of Electronics, vol. 34, pp. 177-183, 1973.

8. S. Kak, The discrete finite Hilbert transform. Indian Journal Pure and Applied Mathematics, vol. 8, pp. 1385-1390, 1977.

9. R. Kandregula, The basic discrete Hilbert transform with an information hiding application, 2009. arXiv:0907.4176

10. R.Kandregula, A DHT Based Measure of Randomness, 2009, arXiv:0908.3689